All-photonic W-band terahertz receiver based on THz-to-optical carrier conversion

with soliton microcomb dual carriers for high-speed OOK wireless transmission

Yudai Matsumura[1,†], Hiroki Kishikawa[2,3,†], Naoya Kuse[2,3], Yasuhiro Okamura[4], Eiji

Hase[2,3], Jun-ichi Fujikata[2,3], Masanobu Haraguchi[2,3], Takahiro Kaji[5], Akira Otomo[5],

Isao Morohashi[5], Tomohiro Tetsumoto[5], Shintaro Hisatake[6], Atsushi Kanno[7], Yu

Tokizane[2,3,8], and Takeshi Yasui[2,3,9]

[1]Graduate School of Sciences and Technology for Innovation, Tokushima University,

2-1, Minami-Josanjima, Tokushima, Tokushima 770-8506, Japan

[2]Institute of Post-LED Photonics (pLED), Tokushima University, 2-1, Minami-

Josanjima, Tokushima, Tokushima, 770-8506, Japan

[3]Institute of Photonics and Human Health Frontier (IPHF), Tokushima University, 2-1,

Minami-Josanjima, Tokushima, Tokushima, 770-8506, Japan

[4]Center for Higher Education and Digital Transformation, University of Yamanashi, 4-

4-37, Takeda, Kofu, Yamanashi 400-8510, Japan

[5]National Institute of Information and Communications Technology, 4-2-1

Nukuikitamachi, Koganei, Tokyo 184-8795, Japan

[6]Department of Electrical, Electronic and Computer Engineering, Gifu University, 1-1,

Yanagido, Gifu, Gufu 501-1193, Japan

[7]Department of Electrical and Mechanical Engineering, Nagoya Institute of

Technology, Gokiso-cho, Showa, Nagoya, Aichi 466-8555, Japan

[8]tokizane@tokushima-u.ac.jp

[9]yasui.takeshi@tokushima-u.ac.jp

[†]These authors contributed equally to this article.

**Abstract**

We demonstrate an all-photonic terahertz receiver for a data-modulated signal, targeting a 106-GHz, 2.97-Gb/s OOK link. The scheme employs dual-wavelength optical carriers referenced to a soliton microcomb and performs THz-to-optical carrier conversion via nonpolarimetric electro-optic downconversion using an electro-optic polymer modulator. RF spectra and eye diagrams confirmed error-free transmission with a Q-factor of 5.78 and a bit-error rate of $3.73 \times 10^{-9}$, well below the hard-decision forward-error-correction (HD-FEC) threshold (Q = 2.67, BER = $3.8 \times 10^{-3}$). Comparative measurements using a single-wavelength optical-carrier configuration clearly revealed the superior signal-to-noise performance of the dual-wavelength scheme. System-level modeling further indicated scalability of the transmission distance beyond 100 m. These results establish soliton microcomb-referenced dual carriers as a promising platform for compact, integrated receivers enabling seamless wireless–optical convergence in future 6G networks.

## 1. Introduction

Sixth-generation (6G) wireless systems aim to deliver ultrahigh data rates, massive connectivity, and ultralow latency, far beyond the capabilities of current fifth-generation (5G) networks [1]. To achieve these goals, 6G is expected to employ carrier frequencies well beyond 300 GHz, while in the early stages of deployment, more practical bands around 100 GHz such as the W- and D-bands are likely to be used first. In these frequency regions of terahertz (THz) wave, conventional wireless electronics that have supported 5G may encounter fundamental bottlenecks in terms of bandwidth, transmission loss, and phase noise. As a result, photonic technologies are attracting strong interest as promising alternatives [2]. On the generation side, microresonator-based optical frequency combs (microcombs) [3–5] have emerged as powerful photonic sources, because they generate mutually coherent optical carriers across multiple wavelengths, with an optical frequency spacing ($f_{rep}$) beyond 100 GHz equivalent to the THz carrier frequency. When combined with uni-traveling-carrier photodiode (UTC-PD)-based photomixing [6], their high-frequency, low-phase-noise $f_{rep}$ signal enables the generation of low-phase-noise THz waves without the phase-noise degradation associated with frequency multiplication [7–11]. Such photonic THz sources have already been applied to wireless transmission experiments [11-15]. On the detection side, however, most THz wireless link demonstrations still rely on electronic devices such as Schottky barrier diodes and heterodyne mixers. Although technically mature, these detectors are often bulky, complicated, and expensive, which poses a serious obstacle to widespread use. Moreover, from the perspective of realizing the ultralow-latency performance targeted by 6G, it is essential to establish a

- 4 -

seamless interface between wireless and optical domains. In this context, direct carrier conversion from THz to the optical domain, followed by detection with mature photonic components, emerges as a key enabling technology for next-generation all-photonic THz receivers, paving the way for truly seamless wireless-optical integration.

In pursuit of seamless wireless-optical connectivity, several approaches to THz-to-optical carrier conversion have been investigated for wireless communication applications. Early studies employed lithium niobate (LN) electro-optic modulators (LN- MODs), leveraging the maturity and wide availability of this material [16-18]. LN-MOD-based schemes demonstrated the feasibility of converting THz carriers into optical signals and thus provided an important foundation for exploring photonic detection in wireless communication systems. However, these demonstrations suffered from intrinsic limitations: the moderate electro-optic coefficient of LN restricted the conversion efficiency, while material absorption in the sub-THz band and the limited scalability of the device hindered practical implementation. As an alternative to LN-based modulators, electro-optic polymer (EOP) modulators (hereafter referred to as EOP-MODs) have attracted attention owing to their superior frequency response and low THz absorption [19, 20]. Building on these advantages, EOP-MOD-based receiver architectures have recently been explored as promising candidates for efficient THz-to-optical carrier conversion.

A recent EOP-MOD-based THz-to-optical carrier conversion scheme [21] incorporated three key features: (i) efficient THz-optical interaction enabled by the EOP-MOD [19, 20], (ii) dual-wavelength near-infrared (NIR) carriers generated via optical injection locking (OIL) of continuous-wave lasers to an electro-optic modulator optical frequency comb (EOM-OFC, frequency spacing $f_{rep}$ = 10 GHz) [22, 23], and (iii) asynchronous nonpolarimetric electro-optic downconversion (NP-EO-DC) [24, 25], in which modulation sidebands created by the EO effect were enhanced through optical heterodyne detection. With this configuration, an unmodulated 101-GHz THz wave and dual-wavelength optical carriers separated by 100 GHz (= $10f_{rep}$) produced a stable 1-GHz RF beat signal with a 20-dB signal-to-noise ratio. The stability and robustness of this result marked an important step toward practical all-photonic THz receivers, paving the way for seamless wireless-optical integration in future 6G systems. Despite these advances, important challenges remain before THz-to-optical carrier conversion can be fully harnessed for practical wireless communication. Previous demonstrations have been limited to unmodulated THz carriers, serving only as proof-of-concept experiments. The critical step of receiving and demodulating modulated THz signals, such as on–off keying (OOK), has not yet been achieved. Furthermore, broadband operation across multi-gigahertz channels, which is essential for high-capacity links, has not been demonstrated. Addressing these issues requires system-level optimization of the entire receiver architecture to ensure sufficient bandwidth and sensitivity for data-modulated THz signals.

In this work, we move beyond proof-of-concept demonstrations with unmodulated carriers and focus on a practical communication scenario in the W-band. Specifically, we target a modulated THz signal at a carrier frequency of 106 GHz employing 2.97-Gb/s OOK. To provide a compact dual-wavelength optical reference, we utilize a soliton microcomb with a repetition rate of 100 GHz. This configuration enables the generation of mutually coherent dual optical carriers with the required frequency spacing and a high optical signal-to-noise ratio (OSNR) via OIL. The incoming THz wave is downconverted by NP-EO-DC, yielding an RF baseband signal centered at 6 GHz. Finally, we measure the eye diagram of the OOK signal and assess the link performance by calculating the Q-factor and the corresponding bit-error rate (BER), thereby providing a quantitative evaluation of communication quality. Through this demonstration, the present study takes a significant step toward all-photonic THz receivers, showing that photonic carrier conversion can be extended from unmodulated proof-of-concepts to the reception of data-modulated THz signals, and opening a pathway to seamless wireless-optical integration for 6G networks.

## 2. Principle of operation

For clarity, the principle of THz-to-optical carrier conversion is explained in two steps: (i) a single optical carrier is used to illustrate the basic operation of the EOP-MOD, which demonstrates the fundamental concept but suffers from a limited signal-to-noise ratio (SNR) due to the weak modulation sidebands, and (ii) dual optical carriers are employed to demonstrate NP-EO-DC with enhanced SNR via optical heterodyne detection.

2.1 THz-to-optical carrier conversion with a single optical carrier

The principle of THz-to-optical carrier conversion is illustrated using a single-wavelength optical carrier, as shown in Fig. 1(a). When a modulated THz carrier (wireless freq. = $f_{THz} \pm \Delta f_{THz}$) is applied to an EOP-MOD together with the optical carrier (opt. freq. = $\nu_1$), the electro-optic effect (Pockels effect) generates modulation sidebands (opt. freq. = $\nu_1 \pm f_{THz}$) in the optical spectrum. These sidebands appear symmetrically above and below the original optical carrier frequency, separated by the modulation frequency of the THz wave, and carry the information originally encoded on the THz carrier. An optical bandpass filter (BPF) can be used to extract either of the modulation sidebands (opt. freq. = $\nu_1 + f_{THz}$), and when detected by a photodetector (PD), this yields an electrical signal corresponding to the THz modulation. However, the sidebands are significantly weaker than the optical carrier, as indicated by a low carrier-to-sideband ratio (CSR, typically ~40 dB). Consequently, the resulting SNR is insufficient for reliable data reception. Therefore, while this single-carrier configuration demonstrates the fundamental concept of THz-to-optical carrier conversion, it is not adequate for achieving high-quality detection in communication applications.

2.2 Nonpolarimetric electro-optic downconversion with dual-wavelength optical carriers

In the dual-wavelength optical carrier configuration, two optical carriers are prepared such that their frequency separation (= $f_{beat}$) is close to the THz carrier frequency (= $f_{THz}$). Let us denote the lower-frequency carrier as $\nu_1$ and the higher-frequency carrier as $\nu_2$ (= $\nu_1 + f_{beat}$). The $\nu_1$ carrier is injected into the EOP-MOD together with the incoming THz signal, generating a pair of modulation sidebands (opt. freq. = $\nu_1 \pm f_{THz}$) in the optical spectrum in the same manner as illustrated in Fig. 1(a). The $\nu_2$ carrier, on the other hand, bypasses the EOP-MOD and is combined with the $\nu_1$ spectrum and its sidebands at the output. When they interfere, the upper modulation sideband of the $\nu_1$ carrier (opt. freq. = $\nu_1 + f_{THz}$) appears next to the $\nu_2$ carrier in the optical spectrum. By selecting this spectral pair with an optical bandpass filter and detecting it with a photodiode, an RF baseband signal corresponding to the THz modulation is obtained. In this case, optical heterodyne beating not only downconverts the THz information into the RF band but also enhances the converted RF signal power, thereby improving the SNR. To realize efficient optical heterodyne detection, it is essential that the two optical carriers maintain a constant frequency spacing, remain fully phase-locked to each other, and exhibit high OSNR. A key enabler of such dual-carrier generation is OIL of high-power dual-wavelength lasers to a soliton microcomb, which provides the required phase coherence and spectral purity.

### 3. Experimental setup

Figure 2 presents the experimental configuration, which follows the measurement concept introduced in Fig. 1(b). In contrast to our previous study [21], the present setup employs a soliton microcomb (µOFC) as the optical frequency reference and a data-modulated THz wave as the input.

A ring-type SiN microresonator (Iloomina AB; FSR = 100 ± 2 GHz; Q > $10^6$; group velocity dispersion = -80 ± 20 $ps^2$/km at 1550 nm) was used for soliton generation. To suppress coupling drift and enable high-power pumping, the microresonator was permanently coupled to a high-numerical-aperture (high-NA) polarization-maintaining single-mode fiber (PMF) using an optical adhesive [26]. Soliton initiation and stabilization were achieved with the aid of the auxiliary-laser-assisted thermal balance technique [27, 28], in which a pump laser (wavelength = 1550.8427 nm, power = 1000 mW) and an auxiliary laser (wavelength = 1550.8307 nm, power =794 mW) were launched bidirectionally into the microresonator. This approach effectively compensates for the thermo-optic effect, eliminates tuning-speed dependence, and extends the soliton existence range. The generated µOFC was subsequently passed through a band-stop filter to remove residual pump light and then optically amplified using an erbium-doped fiber amplifier. Although these components are omitted from Fig. 2 for simplicity, their configuration and operating conditions are described in detail elsewhere [26].

The µOFC, centered at 1550 nm with a repetition rate of 100 GHz, was generated in the high-NA-PMF-connected SiN microresonator and was used for the optical frequency reference of OIL [29]. Two µOFC lines separated by a $f_{rep}$ of 100 GHz (µOFC-M1 and µOFC-M2) were selected using tunable ultra-narrowband bandpass filters (BPF1 and BPF2; Alnair labs, CVF-300CL-PM-FA, center wavelength = 1550.20 nm and 1549.50 nm, optical passband = 0.05 nm, insertion loss = 5.5 dB). A pair of distributed feedback (DFB) lasers (DFB1, Gooch & Housego, AA1408-193350-100-PM900-FCA-NA, $\nu_1$ = 193.39 THz corresponding to a wavelength of 1550.2 nm; DFB2, Gooch & Housego, AA1408-193350-100-PM900-FCA-NA, $\nu_2$ = 193.49THz corresponding to a wavelength of 1549.4 nm), each delivering 4 mW output power, were operated as slave lasers. µOFC-M1 and µOFC-M2 were injected into the DFB1 and DFB2 via optical circulators (OCs), respectively. OIL was achieved by tuning the DFB drive currents such that $\nu_1$ (or $\nu_2$) was positioned adjacent to µOFC-M1 (or µOFC-M2) within the locking range of the DFB lasers (typically a few hundred MHz). Stable OIL was thus established.

The $\nu_1$ carrier locked to µOFC-M1 was coupled into a W-band non-coplanar patch-antenna-type EOP-MOD [21], while the $\nu_2$ carrier bypassed the EOP-MOD. A pair of polarization controllers (PCs) was used to maximize the polarization-dependent coupling efficiency into EOP-MOD. A THz carrier was OOK-modulated at 2.97 Gb/s by mixing the output of a ×6 frequency multiplier (Virginia Diodes; frequency = 106 GHz, power = 20 dBm, corresponding to 100 mW) with the output of a serial digital interface (SDI) converter (Mini Converter UpDownCross HD, Blackmagicdesign, 2.97 Gb/s), into which a high-definition serial digital interface (HD-SDI) data stream was converted into a digital modulation signal. The frequency multiplier was equipped with a horn antenna (Virginia Diodes, WR-10 Horn, 75-110 GHz, gain = 24.7 dBi at 106 GHz). After free-space propagation over 200 mm, the modulated THz beam was focused onto the patch antenna by a spherical lens (L1: D = 50 mm, f = 100 mm) and two orthogonally arranged cylindrical lenses (CL1: 30 × 50 mm$^2$, f = 96 mm; CL2: 85 × 85 mm$^2$, f = 50 mm). The THz focus formed on the EOP-M had an elliptical spot size with a major axis of 9.2 mm and a minor axis of 4.6 mm, which was well matched to the dimensions of the EOP-MOD patch antenna.

The optical output from the EOP-MOD contained the $\nu_1$ carrier together with two ±$\nu_1$ modulation sidebands, although only the + $\nu_1$ sideband is illustrated in Fig. 2 for clarity. By using an erbium-doped fiber amplifier (EDFA) in combination with a pair of bandpass filters (BPF3 and BPF4), the + $\nu_1$ sideband was selectively extracted with an appropriate OSNR. When +$\nu_1$ sideband was combined with the $\nu_2$ carrier, the $\nu_2$ carrier appeared adjacent to the + $\nu_1$ sideband. This pair was detected by a photodiode (PD; λ = 1550 nm, RF bandwidth = 10 GHz). The recovered RF baseband signal was then analyzed using an oscilloscope (KEYSIGHT, UXR0402AP, 40 GHz bandwidth, 256 GSa/s sampling rate) and an electrical spectrum analyzer (ESA, KEYSIGHT, N9000B, frequency range 9 kHz–26.5 GHz).

## 4. Results

### 4.1 Basic performance of dual optical carriers optical-injection-locked to two adjacent microcomb lines

Prior to the communication experiment, we evaluated the basic performance of the dual optical carriers. First, a soliton µOFC with a mode spacing of 100 GHz was generated. Figure 3(a) shows the corresponding optical spectrum, measured with a resolution bandwidth (RBW) of 0.1 nm, exhibiting the characteristic spectral envelope of a soliton microcomb. Specifically, the spectra follow a sech$^2$-shaped distribution, with maximum intensity at the center and a symmetric, gradual decay toward both sides. The stable generation of a soliton microcomb was also confirmed by monitoring the characteristic soliton step observed in the temporal waveform of the optical intensity (not shown).

Next, two adjacent µOFC lines near 1550 nm, denoted as µOFC-M1 (m = +1) and µOFC-M2 (m = +2), were extracted and injected into DFB1 and DFB2 for OIL, respectively. Figure 3(b) presents the optical spectrum of DFB1 locked to µOFC-M1 (red trace, RBW = 0.1 nm), namely the $\nu_1$ carrier in Fig. 2. The enlarged view around the carrier frequency is shown by the red trace in Fig. 3(c) (RBW = 0.1 nm). A sharp line spectrum with an optical power of 10 dBm and an OSNR of 70 dB was observed. For comparison, the spectrum of µOFC-M1 is shown as the blue trace, where the optical power remained at −20 dBm and the OSNR was limited to 60 dB. Residual components from neighboring modes are also visible owing to the limited spectral resolution of BPF1. Nevertheless, the $\nu_1$ carrier spectrum closely matches that of

µOFC-M1, confirming that OIL functioned effectively without residual neighboring modes. Similar results were also obtained for the $\nu_2$ carrier (data not shown).

4.2 OOK transmission performance

We observed the RF spectra of the optical beat signal between the $\nu_1$ sideband and the $\nu_2$ carrier, as shown in Fig. 4. When an unmodulated THz wave was applied, a sharp spike-like baseband signal appeared at 6 GHz as expected, with an SNR of 37 dB [Fig. 4(a)]. In contrast, when a 2.97 Gb/s OOK THz wave (corresponding to the HD-SDI standard) was applied, the RF spectrum exhibited a broadened spectral distribution extending over ±2.97 GHz together with three prominent spike-like components [Fig. 4(b)], with an SNR of 26 dB. OOK corresponds to a rectangular pulse train, whose frequency spectrum is known to consist of discrete spike components associated with the periodicity of the carrier, along with a sinc-shaped broad spectral distribution arising from the non-periodic components. Therefore, the coexistence of sharp spikes (carrier and periodic components) and a broad pedestal (data modulation spectrum) is a characteristic feature of OOK signals.

We also note that the broadened spectral component was not perfectly symmetric. Possible causes include: (i) DC imbalance in the transmitted bit sequence (i.e., unequal ratio of "1" and "0"), which biases the baseband spectrum; (ii) nonlinear response of the modulator or driver amplifier, leading to asymmetry in sideband amplitudes; (iii) measurement system effects such as the finite resolution bandwidth (RBW = 1 MHz) of the spectrum analyzer or group delay of the filters; and (iv) residual carrier leakage, since OOK does not ideally suppress the carrier completely, leaving spike-like components whose phase relationship can distort one side of the spectrum.

Despite these effects, an SNR of 26 dB was achieved, which is sufficient for reliable observation of the eye diagram.

Finally, we evaluated the eye pattern of the 2.97-Gb/s OOK signal. The HD-FEC limit is defined as a Q-factor of 2.67, corresponding to a BER of $3.8 \times 10^{-3}$. Q-factor and BER are widely used as standard indicators of signal integrity in digital communication systems. Figure 5 presents the eye diagram of the received signal, which corresponds to the OOK-modulated data stream used in the experiment. A clearly open eye pattern was observed, confirming high signal integrity and stable data transmission with negligible intersymbol interference. Importantly, the measured Q-factor of 5.78 is about 2.16 times higher than the HD-FEC limit, and the BER of $3.79 \times 10^{-9}$ is six orders of magnitude below the HD-FEC threshold. Both results therefore confirm that the system provides a large performance margin over the HD-FEC limit, demonstrating error-free transmission and validating the feasibility of our all-photonic THz detection scheme for reliable high-speed OOK data links.

## 5. Discussion

In this section, we discuss the implications of the experimental results and the prospects for further development of the proposed all-photonic THz receiver. First, we experimentally investigated the advantage of using dual-wavelength optical carriers over a single-wavelength configuration in THz-to-optical carrier conversion. For a direct comparison, the same experimental setup shown in Fig. 2 was used, but the $\nu_2$ carrier was blocked so that only the $\nu_1$ carrier was applied to the EOP-MOD, as illustrated in Fig. 1(a). In this configuration, the $+\nu_1$ modulation sideband alone was detected by the photodiode, and the resulting RF spectrum around DC was measured. Figure 6(a) shows the observed RF spectrum (red trace), together with the noise floor (blue trace) for reference. A broadened spectral distribution corresponding to the OOK modulation was observed near DC; however, the SNR was limited to 10 dB, which is 16 dB lower than that obtained with dual optical carriers [see Fig. 4(b)]. In addition, the spectral bandwidth was narrowed to below 1.5 GHz. Such limited SNR and bandwidth were insufficient to support stable OOK data transmission. Indeed, as shown in Fig. 6(b), the corresponding eye diagram was completely closed, indicating severe signal degradation. The measured Q-factor and BER were 3.54 and $1.99 \times 10^{-4}$, respectively, both of which fell well short of the HD-FEC threshold (Q-factor = 2.67, BER = $3.8 \times 10^{-3}$). These results clearly confirm the superiority of the dual-wavelength optical carrier scheme in terms of both RF spectral characteristics and OOK transmission performance.

Second, regarding the choice of the dual-wavelength optical carrier reference, we employed a μOFC instead of an EOM-OFC, which served as the optical reference in our previous study [21]. In addition to enhancing the versatility of the setup, one of the advantages of using a microcomb is its high repetition rate, which can be set to the THz carrier frequency. This property eliminates the phase-noise degradation associated with frequency multiplication, a benefit that has been well highlighted in photonic THz generation [7–11] and wireless transmission [11-15]. One might therefore expect that this improvement in phase-noise performance would also be reflected in the performance of the present photonic THz receiver. However, OOK used in the present work is an amplitude-modulation format that is intrinsically insensitive to carrier phase fluctuations, because symbol decisions depend solely on intensity levels rather than on phase information. Consequently, the OOK modulation format effectively masks the influence of phase noise. Although the THz source in the present setup was limited to OOK modulation, the influence of phase noise on transmission performance would become more evident if wireless transmission could be extended to phase-sensitive multilevel modulation formats such as BPSK, QPSK, or QAM. In practice, however, for data transmission in the 100-GHz band, the phase noise of an EOM-OFC driven by a commercially available microwave synthesizer is typically lower than that of an unstabilized microcomb. Based on the specifications of standard microwave synthesizers and the estimated frequency-multiplication order, the phase-noise level of an EOM-OFC is typically around -80 ~ -90 dBc/Hz at a 10-kHz offset, whereas that of a free-running microcomb has been reported to be approximately -60 ~ -70 dBc/Hz at the same offset [10, 13]. To fully exploit the potential advantages of microcomb referencing in photonic THz detection based on THz-to-optical carrier conversion, it will therefore be essential to implement active stabilization of the microcomb repetition rate [8–10] and/or to employ phase-sensitive multilevel modulation formats operating in the high-frequency band. These extensions constitute the next stage of our ongoing research.

Finally, we assessed the scalability of the proposed scheme to longer transmission distances by means of system-level simulations of OOK transmission. In the present experiment, the transmission distance was limited to 200 mm due to experimental constraints. However, as the eye diagram in Fig. 5 and its corresponding Q-factor or BER exhibited a large margin relative to the HD-FEC limit, this result indicates the feasibility of extending the transmission distance. We simulated the dependence of the Q-factor on transmission distance. In the simulation, another antenna (gain = 24.7 dBi) identical to that attached onto the output of the frequency multiplier is added before the EOP-MOD, instead of the THz lens optical system (L1, CL1, and CL2 in Fig. 2), in order to efficiently collect the transmitted power of the OOK signal. The transmission distance is defined as the distance between the two antennas. As summarized in Fig. 6, the simulations predict that with the current configuration, HD-FEC-limit performance can be maintained up to 7 m. Furthermore, the introduction of commercially available high-gain antennas with gains exceeding 36.7 dBi is expected to extend the transmission distance up to 110 m. Therefore, these results indicate that the proposed scheme is not restricted to short-range proof-of-concept demonstrations but can be realistically scaled to practical link lengths suitable for future 6G wireless–optical seamless communication scenarios. These findings collectively validate the scalability and robustness of the proposed all-photonic THz receiver architecture. The details of the simulation model and parameters are provided in the Appendix.

## 6. Conclusion

In this work, we have demonstrated, for the first time to our knowledge, all-photonic terahertz (THz) reception of a 106-GHz, 2.97-Gb/s OOK signal using dual optical carriers referenced to a soliton microcomb and nonpolarimetric electro-optic downconversion (NP-EO-DC) with an electro-optic polymer modulator (EOP-MOD). Measurements of the RF spectrum, eye diagrams, and Q-factors confirmed error-free transmission with a large margin relative to the hard-decision forward-error-correction (HD-FEC) threshold, validating the feasibility of high-speed photonic THz detection.

The demonstrated scheme offers key advantages, including compact implementation, frequency-scalable operation, and direct optical referencing to a microcomb that provides dual carriers with precise THz-rate spacing. These features make the architecture well suited for integration into chip-scale receivers and for extension to broader bandwidths or higher-capacity links.

Finally, system-level simulations indicated that the current configuration can sustain error-free performance over transmission distances on the order of meters. The introduction of high-gain antennas is expected to further extend the achievable link distance beyond 100 m. Given its compatibility with photonic integration and compact form factor, the proposed scheme represents a practical step toward seamless wireless-optical convergence for future 6G communication systems.

**Funding.** Ministry of Internal Affairs and Communications of Japan (JPMI240910001, JPJ000254); Cabinet Office, Government of Japan (Promotion of Regional Industries and Universities); Japan Society for the Promotion of Science (JSPS) Program for Forming Japan's Peak Research Universities (J-PEAKS, JPJS00420240022); Tokushima Prefecture, Japan (Creation and Application of Next-Generation Photonics).

**Data availability.** Data underlying the results presented in this paper are not publicly available at this time but may be obtained from the authors upon reasonable request.

## Appendix

<u>System-level simulation for long-distance OOK links</u>

System-level simulations for long-distance transmission were performed using OptiSystem 22.1 (Optiwave Systems Inc.). The simulations were based on the experimental setup shown in Fig. 2, with the received SNR fixed at 26 dB, identical to that measured in the OOK experiment [Fig. 4(b)]. We simulated the propagation characteristics of an OOK-modulated THz signal (frequency = 106 ± 2.97 GHz; power = 20 dBm) to evaluate distance scalability.

Free-space transmission was modeled using the Friis equation, explicitly accounting for path loss, the far-field condition (> 0.75 m at 106 GHz with a 24.7 dBi antenna), and atmospheric attenuation of $5.35 \times 10^{-4}$ dB/m at 106 GHz [31, 32]. A horn antenna with a gain of 24.7 dBi at 106 GHz was assumed at both the transmitter and receiver sides. In addition, high-gain antennas exceeding 36.7 dBi (far-field condition > 34.2 m) were considered at the transmitter side for extended-range scenarios. The signal quality was evaluated by calculating the Q-factor using OptiSystem's standard OOK demodulation library.

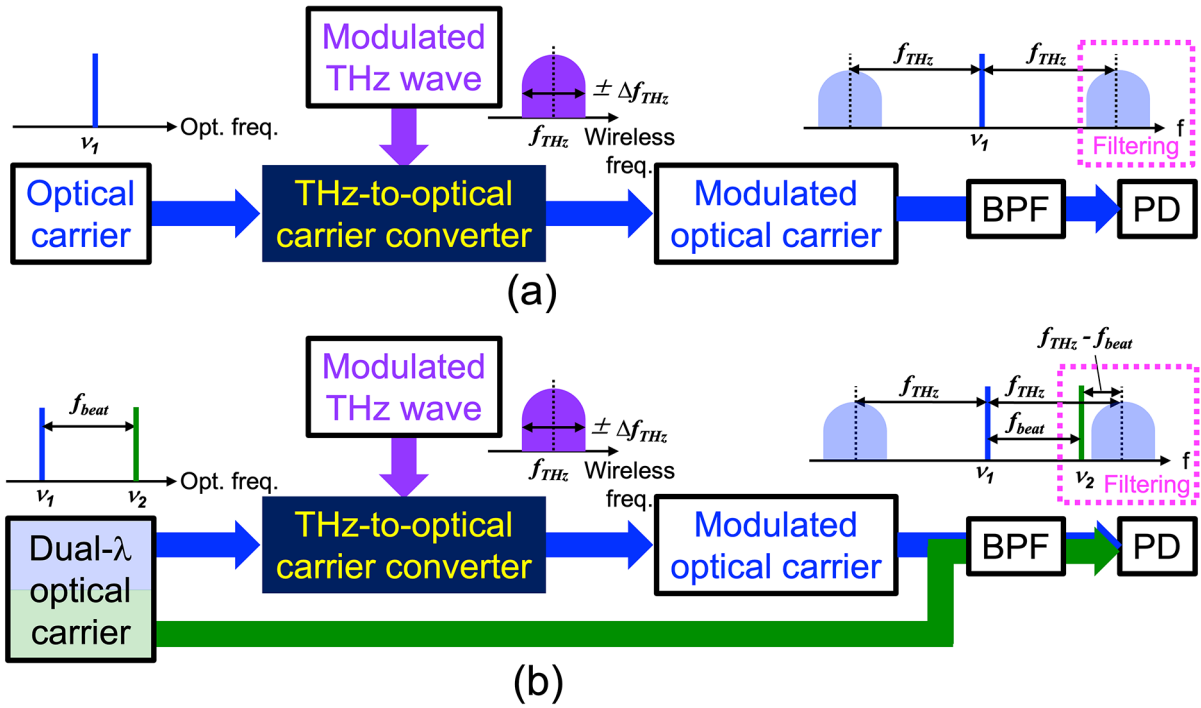

Fig. 1. Principle of THz-to-optical carrier conversion. (a) Basic operation using a single optical carrier (opt. freq. = $\nu_1$). (b) Enhanced operation using dual optical carriers (opt. freq. = $\nu_1$ and $\nu_2$). BPFs, optical bandpass filter; PDs, photodiodes.

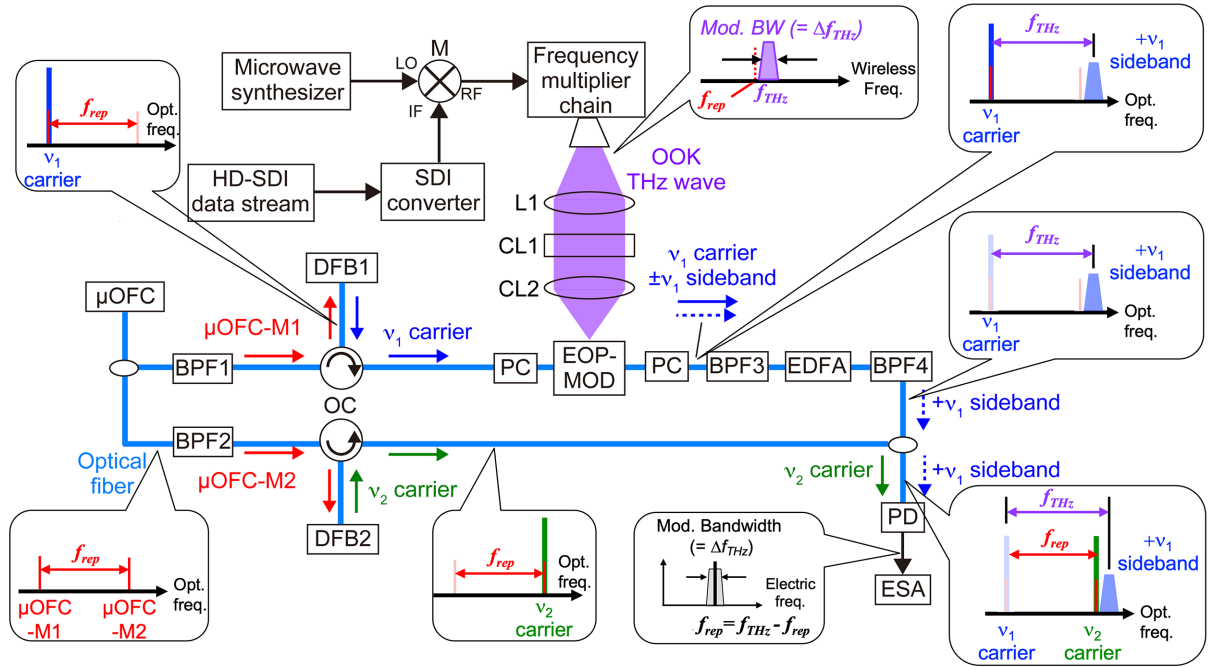

Fig. 2. Experimental setup. μOFC, microresonator-based optical frequency comb; BPF1, BPF2, BPF3, and BPF4, tunable ultra-narrowband optical bandpass filters; OCs, optical circulators; DFB1 and DFB2, distributed feedback laser diodes; PC, polarization controllers; EOP-MOD, electro-optic polymer modulator; EDFA, erbium-doped fiber amplifier; PD, photodiode; ESA, electric spectrum analyzer; L1, spherical THz lens; CL1 and CL2, a crossed pair of cylindrical THz lenses.

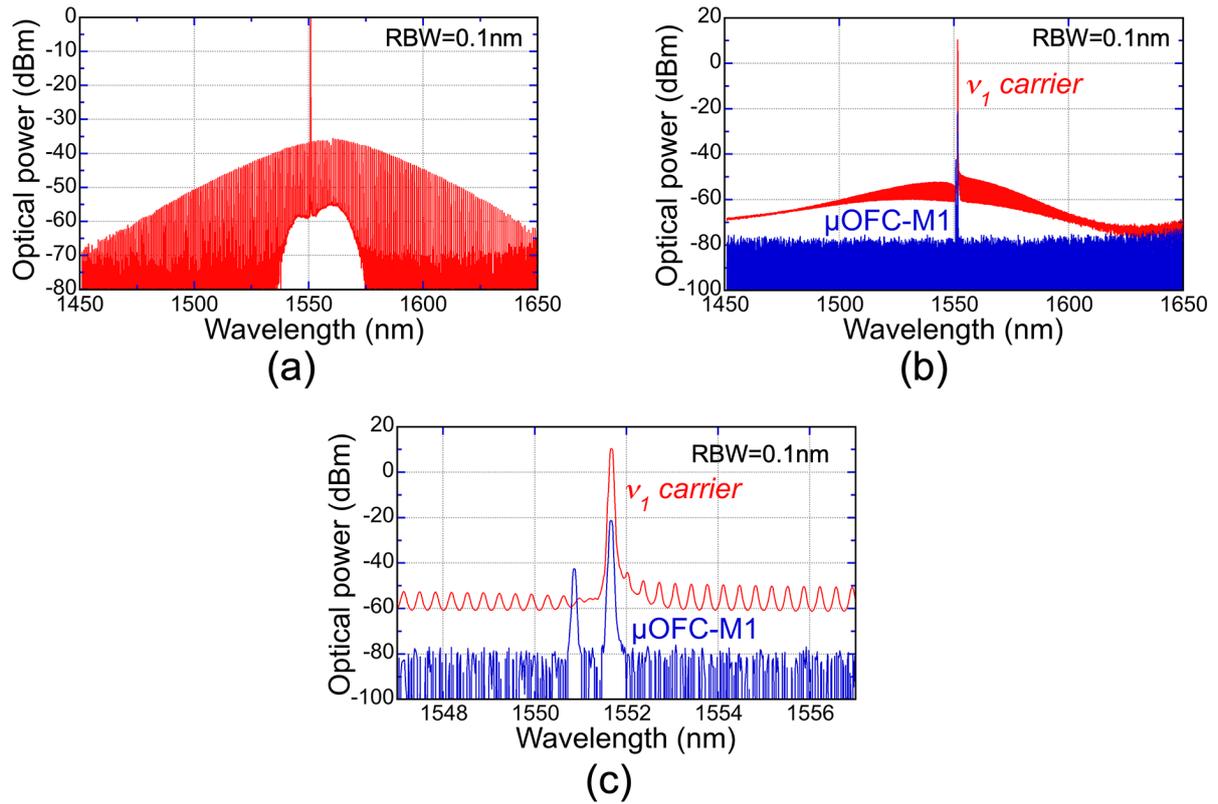

Fig. 3. Basic performance of dual optical carriers optical-injection-locked (OIL) to two adjacent microcomb lines (resolution bandwidth, RBW = 0.1 nm). (a) Optical spectrum of the soliton microcomb with a mode spacing of 100 GHz. (b) Optical spectrum of the DFB1 output ($\nu_1$ carrier, red trace) locked to µOFC-M1, compared with the original µOFC-M1 line (blue trace). (c) Enlarged view of the $\nu_1$ carrier and µOFC-M1 spectra around the locking frequency.

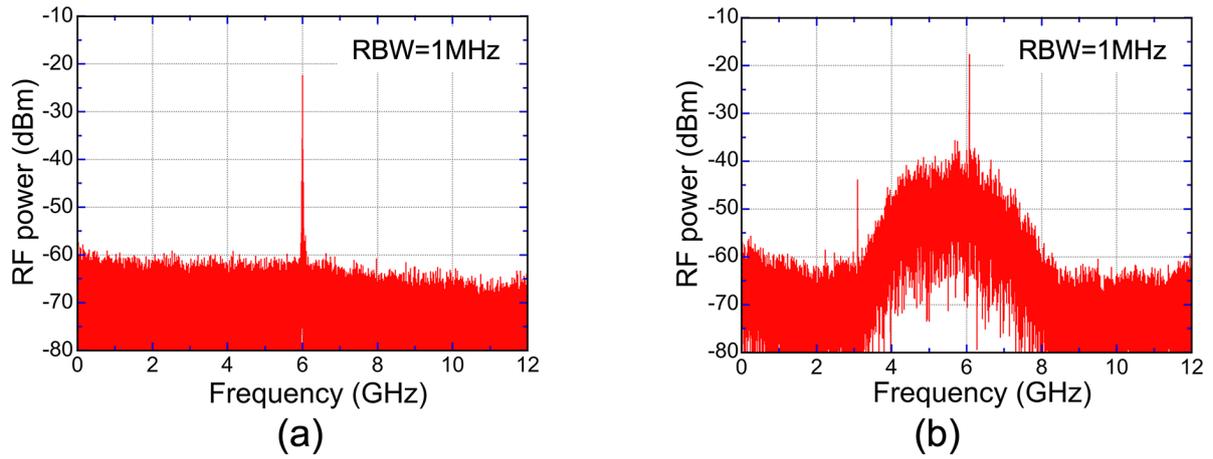

Fig. 4. RF spectra of the optical beat signal between the $\nu_1$ sideband and the $\nu_2$ carrier (resolution bandwidth, RBW = 1 MHz). (a) Sharp spike-like baseband signal when an unmodulated THz wave (wireless freq. = $f_{THz}$ = 106 GHz). (b) Broadened RF baseband spectrum when a 2.97-Gb/s OOK-modulated THz wave (HD-SDI standard) was applied.

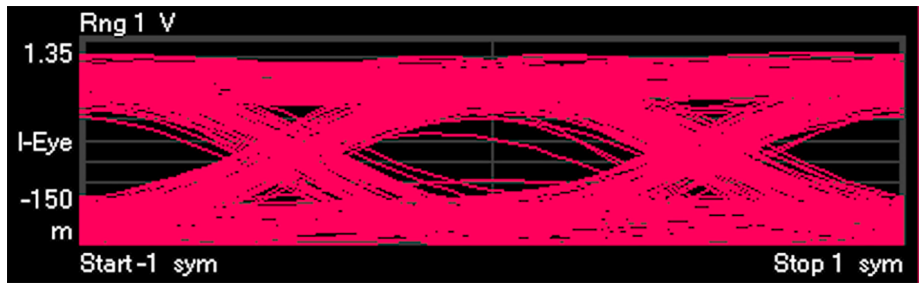

Fig. 5. Eye diagrams of the 2.97-Gb/s OOK signal. The corresponding Q-factor and BER are 5.32 and $5.20 \times 10^{-8}$, respectively.　Both values satisfy the HD-FEC limit (Q-factor = 2.67, BER = $3.8 \times 10^{-3}$).

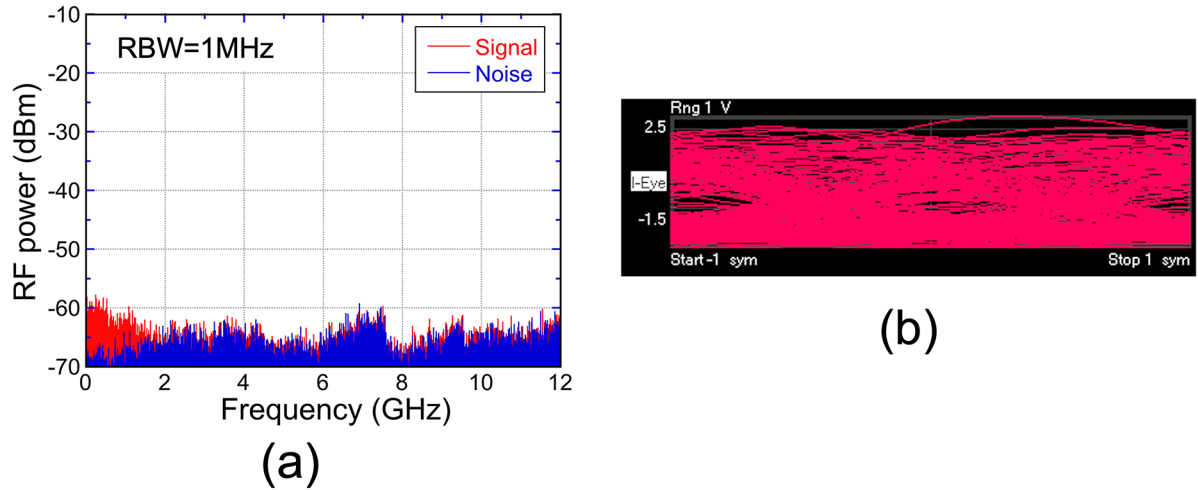

(a)

(b)

Fig. 6. Experimental results of 2.97-Gb/s OOK-modulated THz wave (HD-SDI standard) transmission using a single-wavelength optical carrier. (a) RF spectra of the DC baseband signal (red trace) and the noise floor (blue trace). A broadened spectrum appears near DC with a noticeably reduced SNR (<10 dB) and a narrow spectral bandwidth (<1.5 GHz). (b) Eye diagram of the received OOK signal, showing a completely closed eye. The corresponding Q-factor and BER are 3.54 and $1.99 \times 10^{-4}$, respectively, both of which fail to meet the HD-FEC limit (Q-factor = 2.67, BER = $3.8 \times 10^{-3}$).

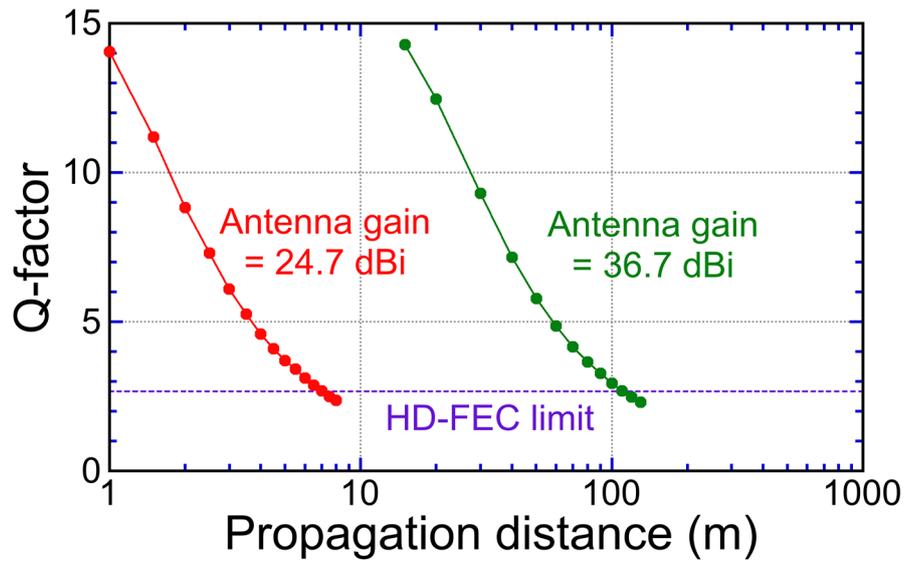

Fig. 7. Simulated dependence of the Q-factor on propagation distance in OOK transmission. The present experimental condition was limited to a 200-mm transmission distance. Simulations predict that with the current configuration, the Q-factor remains above the HD-FEC limit (purple dashed line, Q = 2.67) up to 7 m (red line, antenna gain = 24.7 dBi). The introduction of higher-gain antennas (green line, 36.7 dBi) further extends the feasible range up to 110 m.